\documentclass[aps,pra]{revtex4}
\usepackage{amssymb}
\usepackage{times}

\newcommand{\trace}{\mathop{\rm Tr}\nolimits}

\newcommand{\diag}{\mathop{\rm Diag}\nolimits}

\newcommand{\bra}[1]{\langle#1|}
\newcommand{\ket}[1]{|#1\rangle}
\newcommand{\inpr}[2]{\langle#1|#2\rangle}
\newcommand{\schatten}[2]{\left|\left|\,{#2}\,\right|\right|_{#1}}

\newcommand{\twomat}[4]{{\left(\begin{array}{cc}#1&#2\\#3&#4\end{array}\right)}}

\newcommand{\twovec}[2]{\left(\begin{array}{c}#1\\#2\end{array}\right)}
\newcommand{\twovect}[2]{\left(\begin{array}{cc}#1&#2\end{array}\right)}

\newcommand{\cH}{{\cal H}}

\newcommand{\C}{{\mathbb{C}}}

\newcommand{\id}{\mathrm{\openone}}
\newcommand{\identity}{\mathrm{\openone}}

\newcommand{\be}{\begin{equation}}
\newcommand{\ee}{\end{equation}}
\newcommand{\bea}{\begin{eqnarray}}
\newcommand{\eea}{\end{eqnarray}}
\newcommand{\beas}{\begin{eqnarray*}}
\newcommand{\eeas}{\end{eqnarray*}}

\newcommand{\qed}{\hfill$\square$\par\vskip24pt} 

\newtheorem{theorem}{Theorem}

\newtheorem{corollary}{Corollary}

\newtheorem{proposition}{Proposition}

\begin{document}
\title{Notes on multiplicativity of maximal output purity for completely positive qubit maps}
\author{Koenraad M.R.~Audenaert}
\email{k.audenaert@imperial.ac.uk}
\affiliation{Institute for Mathematical Sciences, Imperial College London,
53 Prince's Gate, London SW7 2PG, UK}
\affiliation{Dept.\ of Mathematics, Royal Holloway, University of London, Egham, Surrey TW20 0EX, UK}
\date{\today}

\begin{abstract}
A problem in quantum information theory that has received
considerable attention in recent years is the question of multiplicativity
of the so-called maximal output purity (MOP) of a quantum channel.
This quantity is defined as the maximum value of the purity one can get
at the output of a channel by varying over all physical input states,
when purity is measured by the Schatten $q$-norm, and is denoted by
$\nu_q$. The multiplicativity problem is the question whether two
channels used in parallel have a combined $\nu_q$ that is the product of the
$\nu_q$ of the two channels. A positive answer would imply
a number of other additivity results in QIT.

Very recently, P.\ Hayden has found counterexamples for every value of $q>1$.
Nevertheless, these counterexamples require that
the dimension of these channels increases with $1-q$ and therefore do not rule out
multiplicativity for $q$ in intervals $[1,q_0)$ with $q_0$ depending on the channel dimension.
I argue that this would be enough to prove additivity of entanglement of formation
and of the classical capacity of quantum channels.

More importantly, no counterexamples have as yet been found
in the important special case where one of the channels
is a qubit-channel, i.e.\ its input states are 2-dimensional.
In this paper I focus attention to this qubit case
and I rephrase the multiplicativity conjecture
in the language of block matrices and prove the conjecture in a number
of special cases.
\end{abstract}

\maketitle
\section{Introduction\label{sec:intro}}
Additivity problems are amongst the most important, and notorious, open problems of quantum information theory.
Basically, the question is whether or not certain information theoretic properties of composite quantum systems consisting
of independent parts decompose as simple sums over these parts.

One of the more important instances of this question concerns the classical information carrying capacity
of quantum channels. Is the total capacity of two quantum channels taken in parallel equal to the sum of the capacities
of the separate channels? Roughly speaking, the classical capacity of a quantum channel quantifies the maximal achievable rate
of error-free communication of classical information through the channel, when the classical information is encoded onto
quantum states that are subsequently transmitted through the quantum channel and then decoded into classical information again \cite{holevo99}.
By judiciously choosing encoding and decoding, the transmission errors incurred when passing through the quantum channel can be corrected.
Theoretically speaking, for every channel, error correcting block codes can be devised so that the remaining probability of error
vanishes asymptotically, when block size goes to infinity. The rate of a code is the number of classical bits of information carried,
on average, by one quantum bit (qubit).
The capacity of the channel is then the maximal rate of an error correcting code that asymptotically corrects all errors for that
particular channel.

A basic result of classical information theory is that the capacity of two classical channels in parallel is just the sum of the
two capacities.
The additivity question for the classical capacity of a quantum channel asks whether this is still true for quantum channels
with encoding/decoding based on quantum error correcting codes. If not, this means that the rate of transmission through the
two channels can be increased by encoding the two streams of classical bits into one stream of \textit{entangled} quantum states,
rather than into two independent streams of quantum states.

Other additivity questions in QIT concern the \textit{entanglement of formation} of bipartite quantum states, which is an important
measure of entanglement, and the \textit{minimal output entropy} of a quantum channel.
A surprising result of quantum information theory is that all these additivity questions are in fact equivalent \cite{ka03,shor},
despite the seemingly different contexts in which they are formulated.
In this paper, I concentrate on what looks like the simplest instance of the additivity questions, namely the additivity of
the minimal output entropy of a quantum channel.

When a pure state is sent through a quantum channel, i.e.\ when a quantum operation acts on a pure quantum state, the resulting
state will in general be no longer pure but will be mixed. By expressing the purity of the resulting state in terms of a
mathematical measure of purity, one can ask for the largest possible value of purity an output state can have when
one can choose the input state freely. One such measure of purity is the von Neumann entropy $S(\rho):=-\trace[\rho\log\rho]$.
As this is an inverted measure of purity (0 for pure states, positive for mixed states), this has to be minimised, yielding the \textit{minimal
output entropy} of a quantum channel. Its precise definition is
$$
\nu_S(\Phi) := \min_\psi S(\Phi(\ket{\psi}\bra{\psi})),
$$
where $\Phi(\cdot)$ denotes the action of the quantum channel on a state. As is well-known, a quantum channel is
mathematically defined as a trace-preserving completely positive map.

A quantity that is closely related to the minimal output entropy is the \textit{maximal output $q$-purity} (MOP). Here, purity is measured
by the Schatten $q$-norm
$$
||\rho||_q := \trace[\rho^q]^{1/q},
$$
a non-commutative generalisation of the familiar $\ell_q$ vector norm.
This yields for the MOP:
$$
\nu_q(\Phi) := \max_\psi ||\Phi(\ket{\psi}\bra{\psi})||_q.
$$
The entropy is related to the Schatten norms via the limit
$$
-x\log x = \lim_{q\to 1} \frac{1-x^q}{q-1}.
$$
In \cite{ahw}, it was proven that the minimal output entropy is additive if the maximal output $q$-norm is multiplicative for
all values of $q$ ``close to 1''; more precisely, if for a pair of given channels $\Phi$ and $\Omega$ there exists a number $q_0>1$ such that
for all $1\le q< q_0$,
$$
\nu_q(\Phi\otimes\Omega) = \nu_q(\Phi)\,\,\nu_q(\Omega)
$$
holds.

Most of the recent efforts on additivity has been directed towards this multiplicativity problem, because of its simple formulation
(and because of the wealth of available techniques for dealing with Schatten norms).
Indeed, when comparing the multiplicativity problem to the other additivity problems, this one almost looks too simple.
However, closer investigation of the equivalence theorems reveal that the complexity is hidden in the dimension
of the channel. More precisely, Theorem 2 from \cite{ka03} states that ``if there exists a real number $q_0>1$ such that
$\nu_q(\Lambda)$ is multiplicative for all $1\le q\le q_0$
and for any CP map $\Lambda$,
then the entanglement of formation is strongly superadditive'', while according to the main result in \cite{msw}, strong superadditivity
of the entanglement of formation implies additivity of the classical capacity of quantum channels.
These two theorems do not mention dimensionality of the states/CP maps/channels involved, because they are stated, in a global manner, in terms of
the set of \textit{all} states/CP maps/channels.
However, on closer inspection of the proofs one finds that something more specific has actually been proven, in terms of the sets of all
CP maps/channels with specified input and output dimension (note, however, that Shor's equivalence theorems do not offer this possibility):
multiplicativity of $\nu_q$, with $q\downarrow 1$, for all CP maps of dimension
\beas
\Lambda_1:&& \cH_{1,in}\mapsto\cH_{1,out} \\
\Lambda_2:&& \cH_{2,in}\mapsto\cH_{2,out},
\eeas
implies
additivity of classical capacity for all channels of dimension
\beas
\Phi_I:   && \cH_{I,in}\mapsto\cH_{I,out} \\
\Phi_{II}:&& \cH_{II,in}\mapsto\cH_{II,out},
\eeas
with
\beas
\cH_{1,in}  &=& {\cH}_{I,out}^{\otimes 2} \otimes \cH_{I,in} \\
\cH_{1,out} &=& \cH_{I,out} \\
\cH_{2,in}  &=& {\cH}_{II,out}^{\otimes 2} \otimes \cH_{II,in} \\
\cH_{2,out} &=& \cH_{II,out}.
\eeas
As an important example, to prove additivity of the classical capacity of a pair of channels,
where one channel is a qubit channel ($2\mapsto 2$),
one needs to prove multiplicativity of MOP for all pairs of CP maps,
where the first one is of dimension $8\mapsto 2$.
Hence, indeed, as regards the multiplicativity of MOP, the complexity of the classical capacity has been hidden in
the increased input dimension of the channels that have to be considered.

\bigskip

Originally, the hope was that $q_0$ in the statement of the multiplicativity question
could be taken to be infinity. Soon after appearance of \cite{ahw}, however, a counterexample
was found for $q>4.78$, involving two identical channels of dimension $3\mapsto 3$ \cite{hw}.
Very recently, the existence of channels was discovered (in a non-constructive way) that violate multiplicativity
for $q$ arbitrarily close to 1 \cite{hayden,winter}. Note, however, that this does not (yet)
disprove additivity of minimal output entropy because the dimension of the channels involved increases
with the minimal value of $q$ for which they violate multiplicativity.
For fixed dimension, there is always room for multiplicativity for $q$ closer to 1, and by the dimension argument mentioned above
this is all one needs. Thus, the claim mentioned in the title of \cite{hayden} that
``the maximal $p$-norm multiplicativity conjecture is false'' is not entirely correct.

In any case, these counterexamples show that even if multiplicativity holds, proving it in some neighbourhood of 1 will be very hard;
most, if not all, known results on Schatten $q$-norms hold over intervals for $q$ like $[1,\infty)$ or $[1,2]$ and not on such intervals
as $[1,q_0]$ with $q_0$ dimension dependent.

On the other hand, the more interesting channels are the lower-dimensional ones, esp.\ the qubit channels, and by the above-mentioned
dimension argument, one can restrict attention to multiplicativity for channels with equally low output dimension.
For qubit channels, no counterexamples have yet been found. In fact,
multiplicativity of MOP when one of the channels is a $2\mapsto 2$ channel has been proven for $q=2$ and $q\ge 4$ \cite{king_p4},
and for all $q\ge1$ when one of the channels is a unital $2\mapsto 2$ channel \cite{king_unital}. Other positive results include multiplicativity
for all $q\ge1$ and for
all dimensions \cite{king} when one of the channels is entanglement breaking (EB) \cite{holevo99},
i.e.\ is of the form $\Phi(\rho) = \sum_k A_k \trace[B_k\rho]$, for $A_k,B_k\ge0$
(that is, the Choi matrix of $\Phi$ corresponds to a separable state).

In this paper I study the multiplicativity problem for the important case when one of the channels has input dimension 2,
and reduce the problem to a number of simpler forms, some of which do not hold in general but can be proven
in specific instances. While the results I obtain here do not boil down to new multiplicativity results, I
do explore new mathematical methods, and the hope is that this will provide new inspiration to tackle the additivity problem.
\section{Notations\label{sec:not}}
In this paper, I call a qubit map any linear map from $\C^2$ to $\C^d$, $d\ge 2$.
This is more general than the customary definition, by which $d=2$.
The reason for this deviation is that the Theorems and Conjectures extend naturally
to these generalised qubit maps.

I will employ overloaded notation where the symbol $\Phi$ either refers to the map or to the Choi matrix of that map.
For example, in expressions like $\Phi(\rho)$, $\Phi$ refers to the map; when used ``stand-alone'',
as in $||\Phi||_q$, it refers to the Choi matrix.

I denote the blocks of the Choi matrix of $\Phi$ by $\Phi_{ij}:=\Phi(e^{ij})$.

Unitarily invariant (UI) matrix norms are denoted $|||.|||$ and are norms that have
the property $|||UAV|||=|||A|||$ for any unitary $U$ and $V$.
For such norms the equality $|||AA^*|||=|||A^*A|||$ holds. This follows from the inequality $|||AB|||\le|||BA|||$
which holds for all $A$ and $B$ such that $AB$ is normal (\cite{bhatia}, Proposition IX.1.1).
When $B=A^*$, both $AB$ and $BA$ are normal, hence equality must then hold.
\section{A Conjecture for Qubit CP Maps\label{sec:conj}}
Let $\Phi$ be a CP qubit map from $\C^2$ to $\C^{d_1}$,
and let $\rho$ be a $2\times d_2$ state, block partitioned as
$$
\rho = \twomat{B}{C}{C^*}{D}.
$$

In \cite{kingconj}, C.\ King conjectured, and proved in specific instances, that for $q\ge 1$
\be\label{eq:chris0}
\schatten{q}{(\Phi\otimes\id)(\rho)} \le
\nu_q(\Phi)\,\, (\beta+\delta),
\ee
where $\beta=\schatten{q}{B}$ and $\delta=\schatten{q}{D}$.
He also noted that this Conjecture would imply multiplicativity of MOP when one of the channels
is a qubit channel.
While this Conjecture is already a major simplification of the multiplicativity problem (it involves only one channel),
it is still non-trivial due to the fact that a maximisation occurs in the RHS (in the factor $\nu_q$).
It would clearly be very helpful if the remaining maximisation could be removed in one way or another.
An initially rather promising idea was that
the following inequality would imply (\ref{eq:chris0}) \cite{kingpriv}:
\be\label{eq:case3}
\schatten{q}{(\Phi\otimes\id)(\rho)} \le
\max_\theta \schatten{q}{\Phi(\twomat{\beta}{\exp(i\theta)\sqrt{\beta\delta}}{\exp(-i\theta)\sqrt{\beta\delta}}{\delta})}.
\ee
That is, the maximisation over all pure qubit input states is replaced by a maximisation over a single angle $\theta$.

To see how this implies multiplicativity, note first that the matrix
$$\frac{1}{\beta+\delta}\twomat{\beta}{\exp(i\theta)\sqrt{\beta\delta}}{\exp(-i\theta)\sqrt{\beta\delta}}{\delta}$$
represents a state (in fact, a pure one), so that the RHS of (\ref{eq:case3}) is bounded above by $(\beta+\delta)\nu_q(\Phi)$,
thereby implying (\ref{eq:chris0}).
Now put $\rho=(\id\otimes\Omega)(\tau)$, with $\tau$ a $2\times d$ state,
then the LHS of (\ref{eq:case3}) is $\schatten{q}{(\Phi\otimes\Omega)(\tau)}$.
The block structure of $\rho$ is then given by
$B=\Omega(\tau^{11})$, $D=\Omega(\tau^{22})$,
yielding the inequality $\beta+\delta\le \nu_q(\Omega)(\trace(\tau^{11})+\trace(\tau^{22}))$,
so that the RHS of (\ref{eq:chris0}) is indeed bounded above by $\nu_q(\Omega)\nu_q(\Phi)$, implying multiplicativity of the MOP.

\medskip

Note that, when the off-diagonal block $\Phi_{12}$ (and thus $\Phi_{21}$) is Hermitian,
the optimal value of $\exp(i\theta)$ in the RHS of (\ref{eq:case3}) is $\pm1$. Indeed,
\beas
\Phi(\twomat{\beta}{\exp(i\theta)\sqrt{\beta\delta}}{\exp(-i\theta)\sqrt{\beta\delta}}{\delta})
&=& \beta \Phi_{11}+\delta \Phi_{22} + 2\cos\theta\,\sqrt{\beta\delta}\,\Phi_{12}.
\eeas
This is linear in $\cos\theta$, hence the RHS of (\ref{eq:case3}) is the maximisation of a convex function
(the $q$-norm of the matrix) in $\cos\theta$. As $\cos\theta$ runs over a convex set (the interval $[-1,1]$), the maximum is obtained
in an extreme point, hence $\pm1$.

\medskip

Unfortunately, numerical experiments revealed that (\ref{eq:case3}) does not hold in general;
I will present such a counterexample in the next Section.
Nevertheless, it is the purpose of this paper to study the statement and introduce a number of techniques to prove it in
a variety of special cases.
I start, in the next Section, with the idea of `taking square roots' of CP maps and states.
\section{Taking `Square Roots'\label{sec:sqrt}}
Positivity of $\rho$ and complete positivity of the map $\Phi$ allow us to `take their square roots' and obtain a `square-rooted' version
of inequality (\ref{eq:case3}), in the following sense.
Since $\rho$ is PSD, it can be written as $\rho = X^* X$, where $X$ is a $1\times 2$ block matrix of size
$R\times d_{in}$ ($R$ being the rank of $\rho$).
Denoting $X=(X_1|X_2)$, we have $B=X_1^*X_1$, $D=X_2^*X_2$,
and $C=X_1^*X_2$.

Similarly, $\Phi$ is CP, thus its Choi-matrix can be written as
$\Phi=G^* G$, where $G$ is a $1\times 2$ block matrix with blocks of size $K\times d_{out}$
($K$ is the number of Kraus elements, $d_{out}$ is the dimension of the output Hilbert
space): $G=(G_1|G_2)$.

The LHS of (\ref{eq:chris0}) and (\ref{eq:case3}) is equal to the square of the $2q$-norm of the `square root'
of $(\Phi\otimes\id)(\rho)$:
$$
(\Phi\otimes\id)(\rho) = (G_1\otimes X_1+G_2\otimes X_2)^* (G_1\otimes X_1+G_2\otimes X_2),
$$
so
$$
\schatten{q}{(\Phi\otimes\id)(\rho)} =
\schatten{2q}{G_1\otimes X_1+G_2\otimes X_2}^2.
$$

Likewise, the `square-root' of the RHS of (\ref{eq:chris0}) is
$$
\max_\psi ||\sum_i \psi_i G_i||_{2q}^2\,\,(||X_1||_{2q}^2+||X_2||_{2q}^2)
$$
so that (\ref{eq:chris0}) is equivalent to
\be\label{eq:chris0sqrt}
\schatten{2q}{G_1\otimes X_1+G_2\otimes X_2} \le
\max_\psi \frac{||\sum_i \psi_i G_i||_{2q}}{\sqrt{|\psi_1|^2+|\psi_2|^2}}\,\,\sqrt{||X_1||_{2q}^2+||X_2||_{2q}^2}.
\ee
This says that
$$
\frac{||G_1\otimes X_1+G_2\otimes X_2||_{2q}}{\sqrt{||X_1||_{2q}^2+||X_2||_{2q}^2}}
$$
attains its maximum over all $X_i$ when $X_2=\alpha X_1$, for certain (complex) values of the scalar $\alpha$.

The square-root of the RHS of (\ref{eq:case3}) is
$$
\Phi(\twomat{\beta}{\exp(i\theta)\sqrt{\beta\delta}}{\exp(-i\theta)\sqrt{\beta\delta}}{\delta})
= \left(G_1 \sqrt{\beta} +G_2 \sqrt{\delta} \exp(i\theta)\right)^*
  \left(G_1 \sqrt{\beta} +G_2 \sqrt{\delta} \exp(i\theta)\right),
$$
which can be written as
$$
\schatten{2q}{G_1 ||X_1||_{2q}+G_2 ||X_2||_{2q} e^{i\theta}}^2,
$$
where I used $\beta = ||X_1^* X_1||_q = ||X_1||_{2q}^2$.
In this way, (\ref{eq:case3}) is equivalent with
\be
\schatten{2q}{G_1\otimes X_1+G_2\otimes X_2}
\le \max_\theta \schatten{2q}{G_1 ||X_1||_{2q} + e^{i\theta}G_2 ||X_2||_{2q}}. \label{eq:case3c}
\ee

\medskip

I now present the promised counterexample to inequality (\ref{eq:case3}), in its square-rooted form (\ref{eq:case3c}).
Consider the diagonal matrices
$G_1=X_1=\diag(1,b), G_2=X_2=\diag(b,-1)$, with $0\le b\le 1$;
then the inequality is violated when $2<2q< p_0$, where $p_0$ is a root of
the equation $((1+b)^p+(1-b)^p)(1+b^p) = 2(1+b^2)^p$ in $p$.
Fortunately, this counterexample does not violate multiplicativity since it corresponds to block-diagonal $\rho$ and $\Phi$;
thus $\Phi$ is EB and $\rho$ is separable, whence multiplicativity holds.
\section{Rank One Case}
In this Section, I describe a technique called the method of conjugation, and use it to obtain results
for the cases where either the CP map or the state has rank 1.

The method of conjugation amounts to transforming existing relations into new ones by replacing
the `components' of the expressions by their Hermitian conjugates, and exploiting in one way or another
the fact that for any UI norm $|||AA^*||| = |||A^*A|||$.
What exactly is meant by `components' here very much depends on the situation, and I will describe here
a number of applications to illustrate the method.
This method is not new; it appears, for example, in \cite{bhatia94}.

Suppose we have a $d\times2$ bipartite state $\rho$ in block-matrix form, and we decompose
it as
$$
\rho = \twovec{X^*}{Y^*} \twovect{X}{Y},
$$
then a possible way of conjugating $\rho$ is to conjugate its components $X$ and $Y$.
This gives rise to a new matrix, of different dimensions, which I denote by $\tilde{\rho}$, and which is given by
$$
\tilde{\rho} = \twovec{X}{Y} \twovect{X^*}{Y^*}.
$$
I want to stress here that the tilde is just a label and not a functional operation, quite simply
because that operation is not uniquely defined; infinitely many $X$ and $Y$ exist for one and the same $\rho$, each giving
rise to different $\tilde{\rho}$.

Exactly the same can be done for a $2\mapsto d$ CP map $\Phi$. Let us decompose its Choi matrix as
$$
\Phi = \twovec{G^*}{H^*} \twovect{G}{H},
$$
then conjugation yields the new map
$$
\tilde{\Phi} = \twovec{G}{H} \twovect{G^*}{H^*}.
$$
If $\Phi$ is a map from $\C^2$ to $\C^d$ of rank $R$ (that is, it can be represented by a minimal number of $R$ Kraus
elements) then one can find blocks $G$ and $H$ of size $R\times d$, so that
$\tilde{\Phi}$ is a map from $\C^2$ to $\C^R$ of rank at most $d$.

The relation linking conjugated state and map to their originals is:
for any UI norm
\be\label{eq:conjug}
|||(\tilde{\Phi}\otimes\id)(\tilde{\rho})||| = |||(\Phi\otimes\id)(\rho)|||.
\ee
This is proven by writing the expressions out in terms of the blocks and exploiting $|||AA^*|||=|||A^*A|||$.
Indeed, $(\Phi\otimes\id)(\rho) = (G\otimes X+H\otimes Y)^*(G\otimes X+H\otimes Y)$, and
$(\tilde{\Phi}\otimes\id)(\tilde{\rho}) = (G\otimes X+H\otimes Y)(G\otimes X+H\otimes Y)^*$.

A simple consequence of (\ref{eq:conjug}) is that $\nu_q(\tilde{\Phi}) = \nu_q(\Phi)$. One just applies (\ref{eq:conjug})
for qubit states $\rho$ (the `blocks' of $\rho$ are scalars) and notes that $\tilde{\rho}$ is the complex conjugate
of $\rho$, whence the maximisation over all $\rho$ coincides with the maximisation over all $\tilde{\rho}$.

The concept of a \textit{complementary channel} introduced in \cite{devetak03,holevo05}
is essentially a specific instance of such a conjugated map.
Let a channel $\Phi$ on a space $\cH$ be represented in Stinespring form by
$$
\Phi(\rho) = \trace_{aux}(U(\rho\otimes\omega)U^*),
$$
where $\omega$ is a fixed ancilla state on the space $\cH_{aux}$, and $U$ is a unitary on $\cH\otimes\cH_{aux}$.
The, or rather `a' complementary channel is then defined as a channel with Stinespring form
$$
\Phi'(\rho) = \trace_{\cH}(U(\rho\otimes\omega)U^*).
$$
Again, for a given $\Phi$, $\Phi'$ is not unique as it depends on the choice of $\omega$ and $U$ \cite{holevo05}.

\begin{proposition}
The complementary channel $\Phi'$ defined above is a conjugated map of the complex conjugation of $\Phi$.
\end{proposition}
\textit{Proof.}
Let the ancilla state $\omega$ be the pure state $\ket{0}\bra{0}$.
If $\Phi$ has Kraus representation $\Phi(\rho)=\sum_k A_k \rho A_k^*$ (where the $A_k$ are defined by
$\langle j|A_k|m\rangle = \langle j,k|U|m,0\rangle$), then the complementary channel $\Phi'$ satisfies
the relation
$\langle k|\Phi'(\rho)|j\rangle = \trace[A_k \rho A_j^*]$.

The Choi matrix of $\Phi$ can thus be decomposed as $\Phi = \twovec{G_1^*}{\vdots} \twovect{G_1}{\cdots}$,
with $G_m^*\ket{k} = A_k\ket{m}$.
Likewise, the Choi matrix of $\Phi'$ can be decomposed as $\Phi' = \twovec{{G'}_1^*}{\vdots} \twovect{{G'}_1}{\cdots}$.
By taking $\rho=\ket{m}\bra{l}$, we find
$$
\langle k|\Phi'(\ket{m}\bra{l})|j\rangle = \trace[A_k \ket{m}\bra{l} A_j^*] = \langle l|A_j^* A_k|m\rangle,
$$
while on the other hand
$$
\langle k|\Phi'(\ket{m}\bra{l})|j\rangle = \langle k|{G'}_m^*\,\,G'_l|j\rangle = \overline{\bra{j}{G'}_l^*\,\,{G'}_m\ket{k}}.
$$
We can, therefore, make the identification
$\overline{{G'}_m}\ket{k}  = A_k\ket{m} = G_m^*\ket{k}$, so that, indeed, $G'_m = \overline{G}_m^* = G_m^T$.
\qed

\bigskip

Using this method of conjugation, we can prove three special cases of the (\ref{eq:case3}).

The first special case is when $\Phi$ is the identity map.
In that case the RHS of (\ref{eq:case3}) reduces to $\beta+\delta$, and we get:
\begin{theorem}
For $\rho\ge0$ partitioned as below, and $q\ge 1$,
\be
\schatten{q}{\rho}=\schatten{q}{\twomat{B}{C}{C^*}{D}} \le ||B||_q + ||D||_q.
\ee
\end{theorem}
This is well-known, and rather easy to prove. In fact, it holds not only for the Schatten norms, but for any UI norm,
and not only for $2\times2$ partitionings, but for any symmetric partitioning.

\textit{Proof.}
The general structure of the proof is: conjugate, apply the triangle inequality, then conjugate again.

By positivity of $\rho$, we can write
$$
\rho = \twovec{X^*}{Y^*} \twovect{X}{Y},
$$
where $X$ and $Y$ are general $d\times 2d$ matrices. Then after conjugating the two factors (not the blocks, but the whole matrix),
we can exploit the triangle inequality
to find
\beas
\schatten{q}{\rho} &=& \schatten{q}{\twovec{X^*}{Y^*}\twovect{X}{Y}} \\
&=& \schatten{q}{\twovect{X}{Y}\twovec{X^*}{Y^*}} \\
&=& ||XX^*+YY^*||_q \\
&\le& ||XX^*||_q + ||YY^*||_q \\
&=& ||X^*X||_q + ||Y^*Y||_q \\
&=& ||B||_q+||D||_q.
\eeas
\qed

\bigskip

An elaboration of the previous argument yields the case
of single-element CP maps $\Phi$, that is maps of the form $\Phi(\rho)=A\rho A^*$.
\begin{theorem}
Inequality (\ref{eq:case3}) holds for $q\ge1$, for $\Phi$ a single-element CP map, and for any state $\rho$.
\end{theorem}
\textit{Proof.}
In this case $A$ has 2 columns, say $A_1$ and $A_2$, and the Choi matrix of $\Phi$
is given by the rank-1 matrix
$$
\Phi = \twovec{A_1}{A_2} \twovect{A_1^*}{A_2^*}.
$$
Let again
$$
\rho = \twovec{X^*}{Y^*} \twovect{X}{Y},
$$
then, by the conjugation identity (\ref{eq:conjug}),
\beas
\schatten{q}{(\Phi\otimes\id)(\rho)} \quad = \quad \schatten{q}{(\tilde{\Phi}\otimes\id)(\tilde{\rho})}
&=& \schatten{q}{A_1^*A_1\, XX^* + A_2^*A_2\, YY^* + A_2^*A_1\, Y^*X + A_1^*A_2\, X^*Y},
\eeas
where I exploited the fact that $A_1$ and $A_2$ are vectors, so that the quantities $A_j^*A_k$ are scalars.
We can do the same thing for the RHS, and get the \textit{scalar} quantity
\be\label{eq:ii}
\schatten{q}{\Phi(\twomat{\beta}{\exp(i\theta)\sqrt{\beta\delta}}{\exp(-i\theta)\sqrt{\beta\delta}}{\delta})}
= \left| A_1^*A_1\, \beta + A_2^*A_2\, \delta
+ A_2^*A_1\,\exp(i\theta)\sqrt{\beta\delta} + A_1^*A_2\,\exp(-i\theta)\sqrt{\beta\delta} \right|
\ee
Comparison of LHS and RHS in this form invites the idea of using the triangle inequality again.
\beas
\schatten{q}{(\Phi\otimes\id)(\rho)}
&\le& A_1^*A_1 \schatten{q}{XX^*} + A_2^*A_2 \schatten{q}{YY^*} + |A_2^*A_1| \schatten{q}{Y^*X}
+ |A_1^*A_2| \schatten{q}{X^*Y}.
\eeas
Now we know that $\schatten{q}{XX^*} = \beta$ and $\schatten{q}{YY^*} = \delta$.
Furthermore, by the Cauchy-Schwarz inequality for UI norms,
$\schatten{q}{Y^*X}\le \schatten{q}{X^*X}^{1/2}\schatten{q}{Y^*Y}^{1/2}=\sqrt{\beta\delta}$.
Thus
\beas
\schatten{q}{(\Phi\otimes\id)(\rho)}
&\le& A_1^*A_1 \beta + A_2^*A_2 \delta + |A_2^*A_1| \sqrt{\beta\delta}
+ |A_1^*A_2| \sqrt{\beta\delta}.
\eeas
By taking $\theta$ such that $|A_2^*A_1| = \exp(i\theta) A_2^*A_1$, the last expression coincides
with RHS(\ref{eq:ii}).
\qed

\bigskip

As a third and final special case, we can in a similar fashion prove (\ref{eq:case3}) for any pure input state $\rho$.
\begin{theorem}
Inequality (\ref{eq:case3}) holds for $q\ge1$, for $\Phi$ a CP map, and for pure states $\rho$.
\end{theorem}
\textit{Proof.}
Let
$$
\Phi = \twovec{X^*}{Y^*} \twovect{X}{Y},
$$
and let $\rho=\ket{\psi}\bra{\psi}$, with
$$
\psi = \twovec{\psi_1}{\psi_2}.
$$
Conjugation of $\rho$, via conjugation of its components $\psi_1$ and $\psi_2$, then gives the qubit state
$$
\tilde{\rho} = \twomat{\inpr{\psi_1}{\psi_1}}{\inpr{\psi_1}{\psi_2}}{\inpr{\psi_2}{\psi_1}}{\inpr{\psi_2}{\psi_2}}.
$$
Thus
$$
\schatten{q}{(\Phi\otimes\id)(\rho)} =
\schatten{q}{\inpr{\psi_1}{\psi_1}XX^* + \inpr{\psi_1}{\psi_2}XY^* + \inpr{\psi_2}{\psi_1}YX^* + \inpr{\psi_2}{\psi_2}YY^*}.
$$
Since we're dealing with a pure state, $\beta=\schatten{q}{\psi_1 \psi_1^*}=\trace(\psi_1 \psi_1^*)=\inpr{\psi_1}{\psi_1}$,
and similarly, $\delta=\inpr{\psi_2}{\psi_2}$.
Also, for some $\theta$, $\inpr{\psi_1}{\psi_2} = e^{i\theta} |\inpr{\psi_1}{\psi_2}| = s e^{i\theta} \sqrt{\beta\delta}$,
with $0\le s\le1$ (by the Cauchy-Schwarz inequality).
Then
$$
\schatten{q}{(\Phi\otimes\id)(\rho)} =
\schatten{q}{\beta X{X}^* + s e^{i\theta}\sqrt{\beta\delta}X{Y}^* + s e^{-i\theta}\sqrt{\beta\delta}Y{X}^* + \delta Y{Y}^*}.
$$
Now notice $s e^{i\theta} = p e^{i\theta}+(1-p)(-e^{i\theta})$ for $p=(1+s)/2$.
Thus
\beas
\schatten{q}{(\Phi\otimes\id)(\rho)} &\le&
p\schatten{q}{\beta X{X}^* + e^{i\theta}\sqrt{\beta\delta}X{Y}^* + e^{-i\theta}\sqrt{\beta\delta}Y{X}^* + \delta Y{Y}^*} \\
&& +(1-p)\schatten{q}{\beta X{X}^* - e^{i\theta}\sqrt{\beta\delta}X{Y}^* - e^{-i\theta}\sqrt{\beta\delta}Y{X}^* + \delta Y{Y}^*} \\
&\le& \max_\theta \schatten{q}{\beta X{X}^* + e^{i\theta}\sqrt{\beta\delta}X{Y}^* + e^{-i\theta}\sqrt{\beta\delta}Y{X}^* + \delta Y{Y}^*} \\
&=& \max_\theta \schatten{q}{\tilde{\Phi}(\twomat{\beta}{e^{i\theta}\sqrt{\beta\delta}}{e^{-i\theta}\sqrt{\beta\delta}}{\delta})} \\
&=& \max_\theta \schatten{q}{\Phi(\twomat{\beta}{e^{-i\theta}\sqrt{\beta\delta}}{e^{i\theta}\sqrt{\beta\delta}}{\delta})}.
\eeas
\qed
\section{Positive Off-Diagonal Blocks}
In the case where the off-diagonal block $\Phi_{12}$ is PSD,
a very general Theorem can be proven for linear maps with general
input and output dimensions.

First we need an Araki-Lieb-Thirring (A-L-T) type inequality for general operators, proven in \cite{kaijiss}:
\begin{proposition}\label{prop:LTG}
For general operators $F$ and $H$, and for $q\ge 1$,
\be
\trace|FHF^*|^q \le \trace\left((F^*F)^q \frac{|H|^q+|H^*|^q}{2}\right).
\ee
\end{proposition}

The following Proposition has appeared before as Lemma 2 in \cite{king}, in somewhat different form, for the case
where all matrices involved are PSD. In that form, the proof relied on the A-L-T inequality. Having now the
stronger inequality from Proposition \ref{prop:LTG} at our disposal, we can lift the original Proposition
to the following more general setting:
\begin{proposition}\label{prop:EB1}
For $A_k\ge 0$ and general $B_k$, and any $q\ge 1$,
\be\label{eq:Apos0}
\schatten{q}{\sum_k A_k\otimes B_k} \le \schatten{q}{\sum_k A_k} \, \max_j \schatten{q}{B_j}.
\ee
\end{proposition}
\textit{Proof.}
Proceeding as in the proof of Lemma 2 in \cite{king},
I introduce the following notations (which are possible because the $A_k$ are PSD):
\beas
F   &=& (\sqrt{A_1} \otimes \identity \ldots \sqrt{A_K} \otimes \identity) \\
G   &=& (\sqrt{A_1} \ldots \sqrt{A_K}) \\
H   &=& \bigoplus_k \identity\otimes B_k.
\eeas
I denote by $X_{kk}$ the $k$-th diagonal block of a matrix in the
same partitioning as $H$.
For example, $H_{kk}=\identity\otimes B_k$.

Using these notations, $\sum_k A_k\otimes B_k$ can be written as $FHF^*$.
By Proposition \ref{prop:LTG},
\beas
\schatten{q}{\sum_k A_k\otimes B_k}^q &=& \trace[|FHF^*|^q] \\
&\le& \trace[(F^* F)^q \, (|H|^q+|H^*|^q)]/2 \\
&=& \sum_k \trace[[(F^* F)^q]_{kk} \, (\id \otimes (|B_k|^q+|B_k^*|^q))]/2 \\
&=& \sum_k \trace[[(G^* G)^q]_{kk}] \, \trace[|B_k|^q] \\
&\le& \max_j \trace[|B_j|^q] \, \sum_k \trace[[(G^* G)^q]_{kk}] \\
&=& \max_j \trace[|B_j|^q] \, \trace[(G^* G)^q].
\eeas
Then noting
$$
\trace[(G^* G)^q] =  \trace[(G G^*)^q] =  \trace[(\sum_k A_k)^q],
$$
and taking $q$-th roots yields the Proposition.
\qed

\begin{corollary}\label{cor:EB1}
For $A_k\ge 0$ and general $B_k$, and any $q\ge 1$,
\be\label{eq:Apos}
\schatten{q}{\sum_k A_k\otimes B_k} \le \schatten{q}{\sum_k A_k \schatten{q}{B_k}}.
\ee
\end{corollary}
\textit{Proof.}
Define $A'_k = ||B_k||_q A_k$ and $B'_k = B_k/||B_k||_q$.
Then $||B'_k||_q=1$ and, by (\ref{eq:Apos0}),
\beas
\schatten{q}{\sum_k A_k\otimes B_k} &=& \schatten{q}{\sum_k A'_k\otimes B'_k}
\le \max_j \schatten{q}{B'_j} \, \schatten{q}{\sum_k A'_k}
= \schatten{q}{\sum_k ||B_k||_q A_k}.
\eeas
\qed

In fact, it is easy to see that the Corollary is equivalent to Proposition \ref{prop:EB1}.
Just note that, by positivity of the $A_k$, $\sum_k ||B_k||_q A_k \le \max_j ||B_j||_q \sum_k A_k$.
The same inequality then holds for the $q$-norm.

\bigskip

Using the above machinery, we can now prove:
\begin{theorem}\label{th:case2}
For linear maps $\Phi$ where \textit{all} the blocks $\Phi_{ij}:=\Phi(e^{ij})$ are positive,
and for general block-partitioned operators $X=[X_{ij}]$:
\be\label{eq:case2}
\schatten{q}{(\Phi\otimes\id)(X)} \le
\schatten{q}{\Phi([\schatten{q}{X_{ij}}])}.
\ee
\end{theorem}
\textit{Proof.}
By assumption, all blocks $\Phi_{ij}$ are positive.
Corollary \ref{cor:EB1} therefore yields
$$
\schatten{q}{(\Phi\otimes\id)(X)}
= \schatten{q}{\sum_{i,j} \Phi_{ij}\otimes X_{ij}}
\le \schatten{q}{\phantom{\big|}\sum_{i,j} \schatten{q}{X_{ij}}\,\Phi_{ij}}
= \schatten{q}{\Phi([\schatten{q}{X_{ij}}])}.
$$
\qed

Proposition \ref{prop:EB1} can also be applied in the `square-root' case.
\begin{corollary}
Let $\Phi$ be a CP map of the form
$$
\Phi=\twovec{G_1^*}{G_2^*}\twovect{G_1}{G_2},
$$
where $G_1$ and $G_2$ are PSD up to scalar phase factors $e^{i\theta_i}$.
Then (\ref{eq:case3}) holds for any state $\rho$ and for $1/2\le q$.
\end{corollary}
Note that this case includes values of $q$ where the ``Schatten $q$-norm'' is not a norm at all.

\textit{Proof.}
Let $G_i=e^{i\theta_i} H_i$, with $H_i$ PSD.
Straightforward application of Corollary \ref{cor:EB1} to LHS(\ref{eq:case3c}) yields, for $2q\ge 1$,
\beas
\schatten{2q}{G_1\otimes X_1+G_2\otimes X_2} &\le&
\schatten{2q}{H_1\otimes e^{i\theta_1} X_1+H_2\otimes e^{i\theta_1}X_2} \\
&=& \schatten{2q}{\schatten{2q}{e^{i\theta_1} X_1} H_1+\schatten{2q}{e^{i\theta_2} X_2} H_2} \\
&=& \schatten{2q}{\schatten{2q}{X_1}G_1+e^{i(\theta_1-\theta_2)}\schatten{2q}{X_2}G_2} \\
&\le& \max_\theta\schatten{2q}{\schatten{2q}{X_1} G_1+e^{i\theta}\schatten{2q}{X_2}G_2},
\eeas
which yields (\ref{eq:case3c}), and hence (\ref{eq:case3}), in this case.
\qed

\bigskip

Proposition \ref{prop:EB1} has some further consequences.
\begin{corollary}\label{cor:sep}
For $\Phi$ a CP map, and $\rho$ a separable state,
\be
||(\Phi\otimes\id)(\rho)||_q \le \nu_q(\Phi)\,\, ||\trace_1\rho||_q.
\ee
\end{corollary}
\textit{Proof.}
Since $\rho$ is separable, it can be written in the form
$\rho=\sum_k \sigma_k\otimes B_k$, where all $\sigma_k$ are normalised states,
and all $B_k$ are positive (not necessarily normalised).
As a consequence, $\sum_k B_k = \trace_1 \rho$.
By Proposition \ref{prop:EB1}, we get
\beas
\schatten{q}{(\Phi\otimes\id)(\rho)}
&=& \schatten{q}{\sum_k \Phi(\sigma_k)\otimes B_k} \\
&\le& \max_k \schatten{q}{\Phi(\sigma_k)} \,\, \schatten{q}{\sum_k B_k} \\
&\le& \nu_q(\Phi)\,\, \schatten{q}{\sum_k B_k} \\
&=& \nu_q(\Phi)\,\, \schatten{q}{\trace_1\rho}.
\eeas
\qed

\begin{theorem}[King]
The MOP is multiplicative for any $q$ when at least one of the CP maps involved is EB.
\end{theorem}
\textit{Proof.}
Let $\Omega$ be an EB CP map, and $\Phi$ any other CP map.
Let $\rho=(\id\otimes\Omega)(\tau)$, with $\tau$ a state.
Because $\Omega$ is EB, $\rho$ is (proportional to) a separable state.
By Corollary \ref{cor:sep}, and the fact that $\trace_1\tau$ is a state, we get
\beas
\schatten{q}{(\Phi\otimes\Omega)(\tau)}
&=& \schatten{q}{(\Phi\otimes\id)(\rho)} \\
&\le& \nu_q(\Phi)\,\, \schatten{q}{\trace_1\rho} \\
&=& \nu_q(\Phi)\,\, \schatten{q}{\Omega(\trace_1\tau)} \\
&\le& \nu_q(\Phi)\,\nu_q(\Omega).
\eeas
\qed

\section{Block-Hankel and Block-Toeplitz Matrices}
Gurvits has proven in \cite{gurvits} that a state whose density matrix is block-Hankel is separable.
For a published reference, see Ando \cite{ando04}, who uses the term `super-positivity' for separability.
For $2\times d$ states, that also follows from the semidefinite programming test in \cite{woerdeman} (using the $n=0$ case).

Gurvits has also proven that states with block-Toeplitz density matrices are separable.
This follows from a representation by Ando (\cite{ando04}, just after Theorem 4.9) which says that
such matrices can be decomposed in terms of a PSD matrix-valued measure $dP(\cdot)$ on the
interval $[0,2\pi)$. For the $2\times d$ case this reads
$$
\rho=\twomat{B}{C}{C^*}{B} = \int_0^{2\pi} \twomat{1}{e^{i\theta}}{e^{-i\theta}}{1} \otimes dP(\theta).
$$
One clearly sees that every factor of the tensor product is positive; $\rho$ is therefore a separable state.
Actually, from the proofs of Lemma 4.8 and Theorem 4.9 in \cite{ando04} one can see that an integral is not needed, and,
instead, we can use a finite sum
$$
\twomat{B}{C}{C^*}{B} = \sum_{k=1}^d \twomat{1}{e^{i\theta_k}}{e^{-i\theta_k}}{1} \otimes P_k.
$$

\bigskip

Using these representations, we can prove one more special instance of (\ref{eq:case3}).
\begin{theorem}\label{th:gen1}
For $\rho=\twomat{B}{C}{C^*}{B}\ge0$, and for $\Phi$ any linear map, (\ref{eq:case3}) holds for all $q\ge1$.
\end{theorem}
\textit{Proof.}
According to Ando's representation mentioned in the previous Section,
$\rho$ can be written in the form
$$
\twomat{B}{C}{C^*}{B} = \sum_{k=1}^d \twomat{1}{e^{i\theta_k}}{e^{-i\theta_k}}{1} \otimes P_k,
$$
with $P_k\ge0$.
Applying the map $\Phi\otimes\id$ gives
$$
(\Phi\otimes\id)(\rho) = \sum_k \Phi(\twomat{1}{e^{i\theta_k}}{e^{-i\theta_k}}{1}) \otimes P_k.
$$
Because $\Phi$ need not be CP, the first tensor factor is no longer positive. However, the $P_k$ still are,
allowing us to employ Proposition \ref{prop:EB1},
which gives us
$$
\schatten{q}{(\Phi\otimes\id)(\rho)} \le \max_{\theta} \schatten{q}{\Phi(\twomat{1}{e^{i\theta}}{e^{-i\theta}}{1})} \,\,
\schatten{q}{\sum_k P_k}.
$$
Noticing that the second factor is just $||B||_q=\beta=\delta$ yields (\ref{eq:case3}).
\qed

\bigskip

When $B$ is different from $D$, restrictions have in general to be imposed on $\Phi$; the previous Theorem being
an exception.
For $C=C*$, the RHS of (\ref{eq:case3}) only depends on $X:=\Phi_{11}+\Phi_{22}$ and $Y:=\Phi_{12}+\Phi_{21}$,
so that we are led to maximise the LHS over all maps $\Phi$, keeping $X$ and $Y$ fixed.
Now the LHS is
$$
\schatten{q}{(\Phi\otimes\id)(\rho)} = \schatten{q}{\Phi_{11} \otimes B + \Phi_{22}\otimes D + Y\otimes C}
= \schatten{q}{X \otimes \frac{B+D}{2} + \Delta\otimes \frac{B-D}{2} + Y\otimes C},
$$
where $\Delta:=\Phi_{11} - \Phi_{22}$.
So if $\Phi$ is unconstrained, and $B-D\neq0$, then the LHS could be made arbitrarily large
by letting $\Delta$ become arbitrarily large.
If, however, $\Phi$ is CP, say, then that can no longer happen.
Indeed, by positivity of $\Phi_{11}$ and $\Phi_{22}$,
$X\pm\Delta\ge0$, whence $-X\le \Delta \le X$.
\acknowledgments
This work was supported by The Leverhulme Trust (grant F/07 058/U),
and is part of the QIP-IRC (www.qipirc.org) supported by EPSRC (GR/S82176/0).
I thank Chris King for many stimulating conversations and his suggestion to study (\ref{eq:case3}).

Dedicated with love to young Ewout Audenaert, whose constant calls for attention kept me awake during the course of this work.

\end{document}